\documentclass[conference]{IEEEtran}

\usepackage{amsmath,amssymb,amsfonts}
\usepackage{algorithmic}
\usepackage{graphicx}
\usepackage{textcomp}
\usepackage{xcolor,colortbl}
\usepackage{subcaption}
\usepackage{multirow}
\usepackage{threeparttable}
\usepackage{makecell}
\usepackage{listings} 
\usepackage{hyperref}
\usepackage{adjustbox}
\usepackage{arydshln}
\usepackage{cancel}
\usepackage{multibib}

\title{OpenDCVCs: A PyTorch Open Source Implementation and Performance Evaluation of the DCVC series Video Codecs}

\author{
    \IEEEauthorblockN{Yichi Zhang}
    \IEEEauthorblockA{
        Purdue University \\
        West Lafayette, USA \\
        zhan5096@Purdue.edu
    }
    \and
    \IEEEauthorblockN{Fengqing Zhu}
    \IEEEauthorblockA{
        Purdue University \\
        West Lafayette, USA \\
        zhu0@Purdue.edu
    }
}

\begin{document}

\maketitle

\begin{abstract}
We present \textbf{OpenDCVCs}, an open-source PyTorch implementation designed to advance reproducible research in learned video compression. OpenDCVCs provides unified and training-ready implementations of four representative Deep Contextual Video Compression (DCVC) models—DCVC, DCVC with Temporal Context Modeling (DCVC-TCM), DCVC with Hybrid Entropy Modeling (DCVC-HEM), and DCVC with Diverse Contexts (DCVC-DC). While the DCVC series achieves substantial bitrate reductions over both classical codecs and advanced learned models, previous public code releases have been limited to evaluation codes, presenting significant barriers to reproducibility, benchmarking, and further development. OpenDCVCs bridges this gap by offering a comprehensive, self-contained framework that supports both end-to-end training and evaluation for all included algorithms. The implementation includes detailed documentation, evaluation protocols, and extensive benchmarking results across diverse datasets, providing a transparent and consistent foundation for comparison and extension. All code and experimental tools are publicly available at \url{https://gitlab.com/viper-purdue/opendcvcs}, empowering the community to accelerate research and foster collaboration.
\end{abstract}


\section{Introduction}

Video compression is a cornerstone of today’s multimedia landscape, especially as the proliferation of ultra-high definition (UHD) content puts increasing pressure on storage and transmission systems. Recent advances in deep learning have enabled end-to-end optimization and the learning of complex, data-driven representations~\cite{lu2019dvc,li2021deep,sheng2022temporal,li2022hybrid,li2023neural,li2024neural}.

Among learned video compression methods, the Deep Contextual Video Compression (DCVC) series stands out for its innovative use of conditional coding in the feature domain. Rather than compressing pixel-wise residuals, DCVC models leverage rich context from previously decoded frames to more effectively model conditional entropy~\cite{li2021deep,sheng2022temporal,li2022hybrid,li2023neural,li2024neural}. This design yields substantial bitrate reductions and superior reconstruction quality, outperforming both classical codecs and earlier learned approaches. Despite these empirical successes, shown across DCVC variants like DCVC-TCM~\cite{sheng2022temporal}, DCVC-HEM~\cite{li2022hybrid}, and DCVC-DC~\cite{li2023neural}—reproducibility and further development have been hindered by the lack of open-source, training-ready implementations. In particular, official repositories provide only evaluation code, with no training scripts, limiting the community’s ability to benchmark and extend.

To address this gap, we present \textbf{OpenDCVCs}: a comprehensive, open-source PyTorch implementation that implements the DCVC series. OpenDCVCs includes four representative deep video codecs—DCVC, DCVC-TCM, DCVC-HEM, and DCVC-DC—and provides complete pipelines for both training and evaluation. The implementation is designed for usability, with detailed instructions, reproducible scripts, and extensive benchmarking on standard datasets. We offer thorough performance comparisons—including rate-distortion curves, runtime analysis, and GPU memory usage—to enable fair and transparent evaluation. By making all code and evaluation tools publicly available, we aim to lower barriers to entry and accelerate research in learned video compression.

Compared to existing open-source efforts, OpenDCVCs offers several unique advantages. \textit{CompressAI} is the leading framework for learned image compression, but provides only limited support for video, only covering SSF~\cite{agustsson2020scale}. \textit{OpenDVC}~\cite{yang2020opendvc} provides a PyTorch implementation for DVC~\cite{lu2019dvc} and does not support advanced architectures such as DCVC. Similarly, \textit{PytorchCompression} focuses on DVC and FVC~\cite{hu2021fvc}. More recently, \textit{OpenDMC}~\cite{gao2023opendmc} has integrated a broader set of video compression models, including DVC, SSF, DCVC, and DVC-P~\cite{zhang2021dvc}. However, its training pipeline lacks the progressive training strategies described in the DCVC series and other recent codecs~\cite{lin2020m}, limiting reproducibility and benchmarking.

Our main contributions are summarized as follows:
\begin{itemize}
    \item We introduce OpenDCVCs, an open-source PyTorch implementation supporting the DCVC series, with accessible, training-ready, and extensible implementations for the research community.
    \item We implement and benchmark four advanced codecs—DCVC, DCVC-TCM, DCVC-HEM, and DCVC-DC—providing extensive experimental results and analysis on standard datasets.
    \item We release comprehensive documentation, training pipelines, and evaluation tools to foster transparency and rapid progress in learned video compression.
\end{itemize}
We anticipate that OpenDCVCs will serve as a valuable foundation for future innovations in learned video coding.

\section{Supported Algorithm Library}

\subsection{Deep Contextual Video Compression}
DCVC~\cite{li2021deep} represents a paradigm shift in deep video compression by introducing conditional coding in the feature domain. Unlike traditional approaches that encode pixel-wise residuals, DCVC uses deep neural networks to extract rich, high-dimensional contextual features from previously decoded frames. These features guide both motion compensation and entropy modeling, enabling the model to capture temporal dependencies. As a result, DCVC achieves superior rate-distortion performance and consistently outperforms both classical codecs like x265 and earlier learned compression models. This framework serves as the basis for further advancements.

\subsection{Temporal Context Mining for Learned Video Compression}
DCVC-TCM~\cite{sheng2022temporal} builds upon the DCVC framework by introducing a temporal context mining (TCM) module. While DCVC extracts single-scale context from previous frames, DCVC-TCM propagates feature representations prior to frame reconstruction and hierarchically mines multi-scale temporal contexts. This approach captures finer and more varied temporal information, allowing the model to better handle spatial-temporal dependencies and non-uniform motion. The temporal context re-filling (TCR) strategy further injects these contexts into multiple modules within the network, enhancing overall compression efficiency. Additionally, by removing the spatial auto-regressive entropy model, DCVC-TCM enables faster encoding and decoding without sacrificing performance.

\subsection{Hybrid Spatial-Temporal Entropy Modelling for Neural Video Compression}
DCVC-HEM~\cite{li2022hybrid} advances the DCVC-TCM by proposing a hybrid entropy model that exploits both spatial and temporal dependencies in video data. In addition to multi-scale temporal context mining, DCVC-HEM introduces a latent prior for modeling correlations among latent variables across frames, and a dual spatial prior for modeling spatial dependencies in a parallel fashion. The dual spatial prior extends context coverage and captures cross-channel relationships, substantially improving probability estimation. DCVC-HEM also integrates a multi-granularity quantization mechanism, supporting flexible variable rate coding within a single model.

\subsection{Neural Video Compression with Diverse Contexts}
DCVC-DC~\cite{li2023neural} further extends DCVC-HEM by leveraging even more diverse temporal and spatial contexts. In the temporal domain, it employs a hierarchical quality structure during training, allowing the model to utilize long-term, high-quality references and better exploit dependencies over larger temporal ranges. The group-based offset diversity mechanism predicts multiple motion offsets for different feature groups, which are then fused to robustly handle complex motion. On the spatial side, DCVC-DC introduces a quadtree-based partition scheme for entropy coding, enabling finer-grained and more diverse spatial context modeling while retaining parallel efficiency. Other architectural enhancements, such as depthwise separable convolutions, further boost both efficiency and compression performance.

\section{Key Modifications for Training}

Although the authors of DCVC and its variants released inference code, these frameworks are not trainable out of the box. The absence of training scripts and details, together with several non-differentiable operations and an inference-focused codebase, presents major obstacles to end-to-end optimization. In developing \texttt{OpenDCVCs}, we addressed several critical challenges to enable efficient, reproducible training for all supported DCVC models.

\subsection{Differentiable Model Components}
\label{subsec:mod_model}

A primary challenge stems from non-differentiable operations in the original models, particularly in quantization and entropy modeling:

\begin{itemize}
    \item \textbf{Quantization Relaxation:} The original inference code employs scalar quantization, which blocks gradient flow during backpropagation. To enable gradient-based optimization, we follow prior efforts in learned image compression to replace hard quantization with a mixed quantization strategy~\cite{minnen2020channel} (additive uniform noise in the range $(-0.5, 0.5)$ for entropy parameter estimation and straight through estimator for reconstruction) during training. This approach simulates quantization effects while preserving differentiability.
    \item \textbf{Stable Entropy Modeling:} The original code clips probabilistic parameters (e.g., $\sigma$ in Laplace or Gaussian models) to ensure non-negativity, but such clipping produces zero gradients and destabilizes training. To address this, we reparameterize the scale parameter using a differentiable lower bound, following~\cite{zhang2024theoretical}. Specifically, we transform the scale parameter as
    \[
    \texttt{scales} = \exp(\texttt{softplus}(\texttt{scales} + 2.3) - 2.3)
    \]
    to guarantee $\texttt{scales} > 0.1$, ensuring numerical stability and effective learning.
\end{itemize}

\subsection{Data Augmentation}

Robust data augmentation is critical for generalization in learned video compression. Since the original DCVC paper does not specify augmentation strategies, we incorporate two effective techniques:
\begin{itemize}
    \item \textbf{Random Horizontal/Vertical Flipping}: Widely used in image and video models.
    \item \textbf{Random Frame Shuffling}: Specifically for video, this technique randomizes frame order within a sequence to encourage resilience to various motion dynamics and temporal changes.
\end{itemize}

\subsection{Training Strategy}

Training DCVCs efficiently and robustly requires a carefully designed strategy that addresses both optimization stability and temporal modeling. We adopt a two-stage strategy consisting of progressive pretraining and multi-frame finetuning, enabling both efficient convergence and improved temporal modeling.

\subsubsection{Progressive Pretraining}

Directly optimizing the full rate-distortion loss for all modules can lead to unstable training and degenerate solutions, such as the model bypassing motion estimation. To address this, we employ a progressive pretraining strategy following~\cite{lin2020m,li2021deep}, in which model components are activated and optimized in sequence. During pretraining, we use a three-frame (IPP) structure: the first frame is encoded as an I-frame, the second as a P-frame referencing the I-frame, and the third as a P-frame referencing the preceding P-frame. This setup achieves a balance between temporal modeling and computational efficiency.

The pretraining proceeds through four steps, each governed by a specific loss function. In the first step (\emph{MV Warm-up}), only the motion estimation and motion vector (MV) encoding/decoding modules are trained, optimizing
\[
L_{\text{me}} = \lambda \cdot D(x_t, \tilde{x}_t) + R(\hat{g}_t) + R(\hat{s}_t)
\]
where $x_t$ is the ground-truth frame, $\tilde{x}_t$ is the motion-warped frame, and $R(\cdot)$ denotes bitrate for the motion latent and its hyperprior. 

In the second step (\emph{Reconstruction Training}), the MV modules are frozen, and the model is optimized with
\[
L_{\text{reconstruction}} = \lambda \cdot D(x_t, \hat{x}_t)
\]
where $\hat{x}_t$ is the reconstruction. 

The third step (\emph{Contextual Coding}) introduces bitrate regularization for the contextual latents, with
\[
L_{\text{contextual\_coding}} = \lambda \cdot D(x_t, \hat{x}_t) + R(\hat{y}_t) + R(\hat{z}_t)
\]
where $\hat{y}_t$ and $\hat{z}_t$ are the quantized content latent and its hyperprior. 

Finally, in the \emph{End-to-End Optimization} step, all modules are jointly trained for the complete rate-distortion objective:
\[
L_{\text{all}} = \lambda \cdot D(x_t, \hat{x}_t) + R(\hat{y}_t) + R(\hat{z}_t) + R(\hat{g}_t) + R(\hat{s}_t)
\]
Each step is trained for around 30 epochs using a ReduceLROnPlateau scheduler (reduction factor 0.5, patience 3), with the learning rate reset to $1 \times 10^{-4}$ at the start of each new stage.

\subsubsection{Multi-frame Finetuning}

While progressive pretraining with three-frame (IPP) sequences is efficient, it is limited in modeling long-term dependencies and error propagation seen in actual video coding deployments. To bridge this gap, we perform an additional fine-tuning stage using longer multi-frame sequences (IPP$\cdots$P), with the sequence length determined by available GPU resources. In this stage, the loss is computed and gradients are backpropagated over all frames in the sequence~\cite{lu2020content}, exposing the model to more realistic temporal dynamics. This multi-frame fine-tuning lasts for 10 epochs with an learning rate of $4 \times 10^{-5}$
and helps mitigate temporal error propagation in longer sequences, further improving rate-distortion performance.

This two-stage strategy—progressive pretraining followed by multi-frame finetuning—enables stable, efficient training while ensuring strong temporal modeling in all DCVC-series models implemented in \texttt{OpenDCVCs}.

\section{Benchmarking Results and Analysis}
\subsection{Datasets}
\textbf{Training Data:}  
We use the training partition of the Vimeo-90k septuplet dataset~\cite{xue2019video} as the source of training samples. During training, video sequences are randomly cropped into $256 \times 256$ patches.

\textbf{Testing Data:}  
For testing, we evaluate our models on benchmark datasets widely used in the video compression literature: HEVC Class B~\cite{jvetj1010}; UVG~\cite{uvg2021}; MCL-JCV~\cite{wang2016mcl}.

\textbf{Test Conditions:} 
We test 96 frames for each video, and the intra period is set as 32. The low delay encoding setting is used. 
BD-Rate is used to measure the compression ratio, where negative numbers indicate bitrate saving and positive numbers indicate bitrate increase. Anchor is DCVC with the officially released checkpoints.

\begin{figure*}[htbp] 
\newcommand{\mywidth}{0.3}
\centering 
\begin{subfigure}[b]{\mywidth\linewidth}
    \centering
    \includegraphics[width=\linewidth]{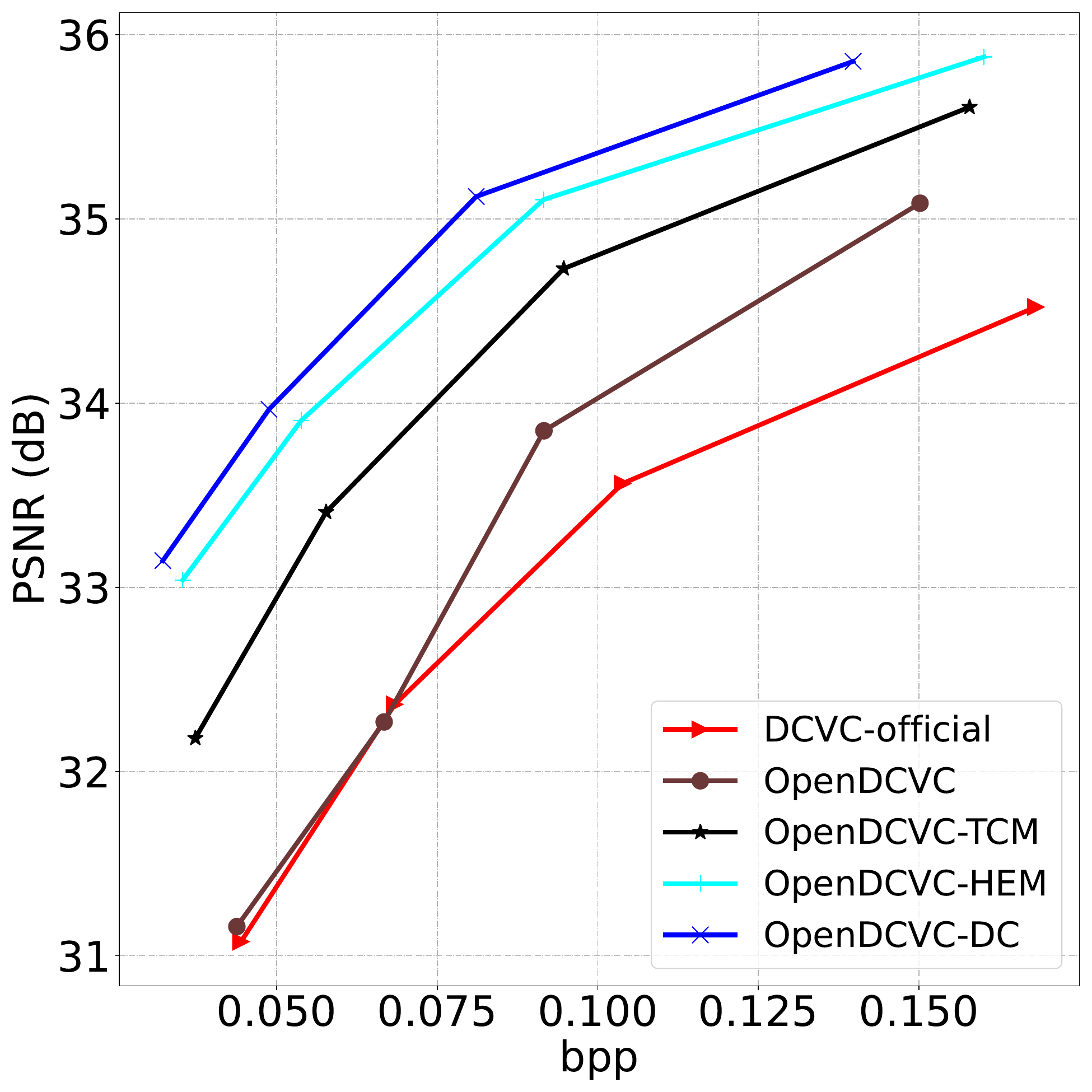}
    \caption{HEVC-B}
    \label{subfig:kodak}
\end{subfigure}
\hfill
\begin{subfigure}[b]{\mywidth\linewidth}
    \centering
    \includegraphics[width=\linewidth]{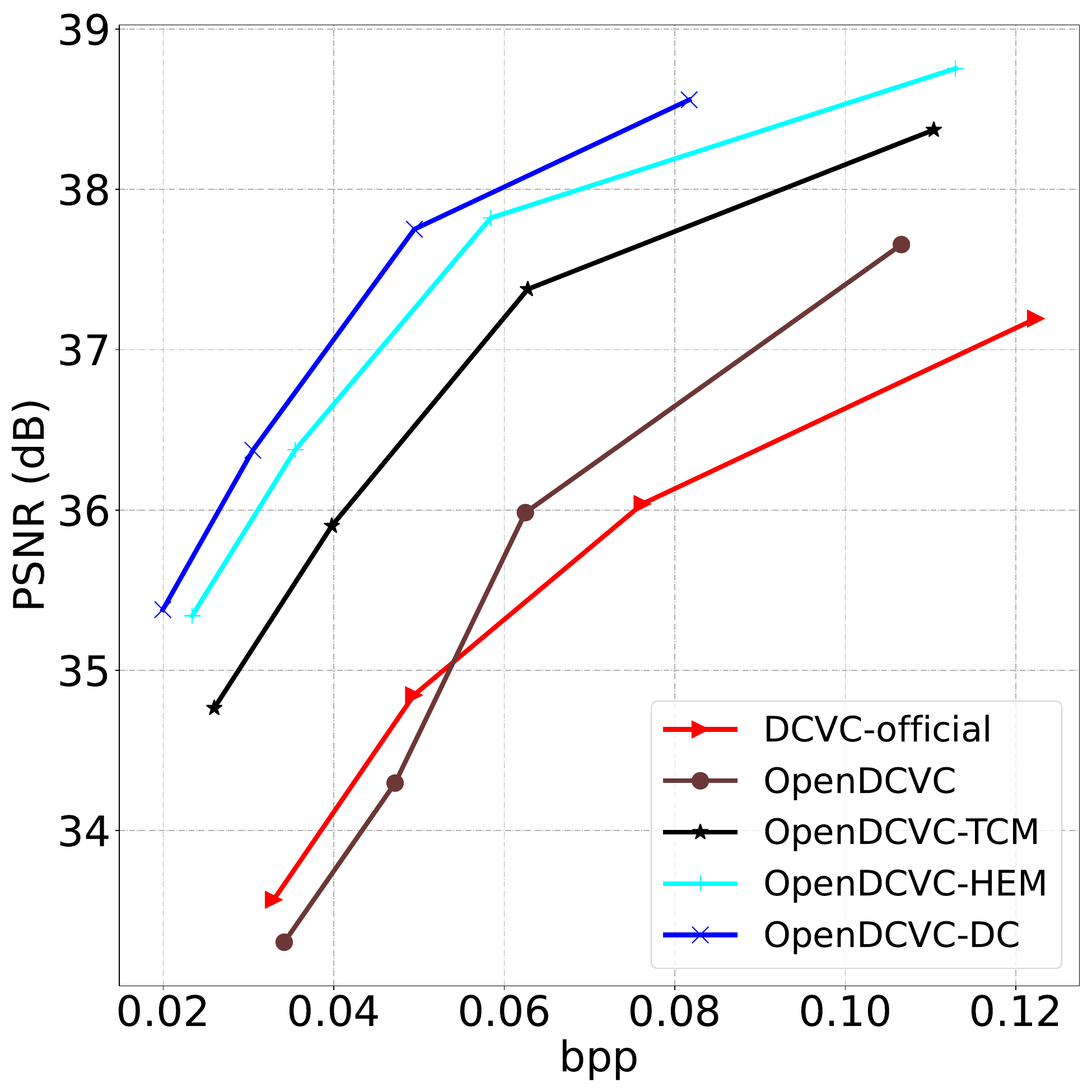}
    \caption{UVG}
    \label{subfig:CLIC}
\end{subfigure}
\hfill
\begin{subfigure}[b]{\mywidth\linewidth}
    \centering
    \includegraphics[width=\linewidth]{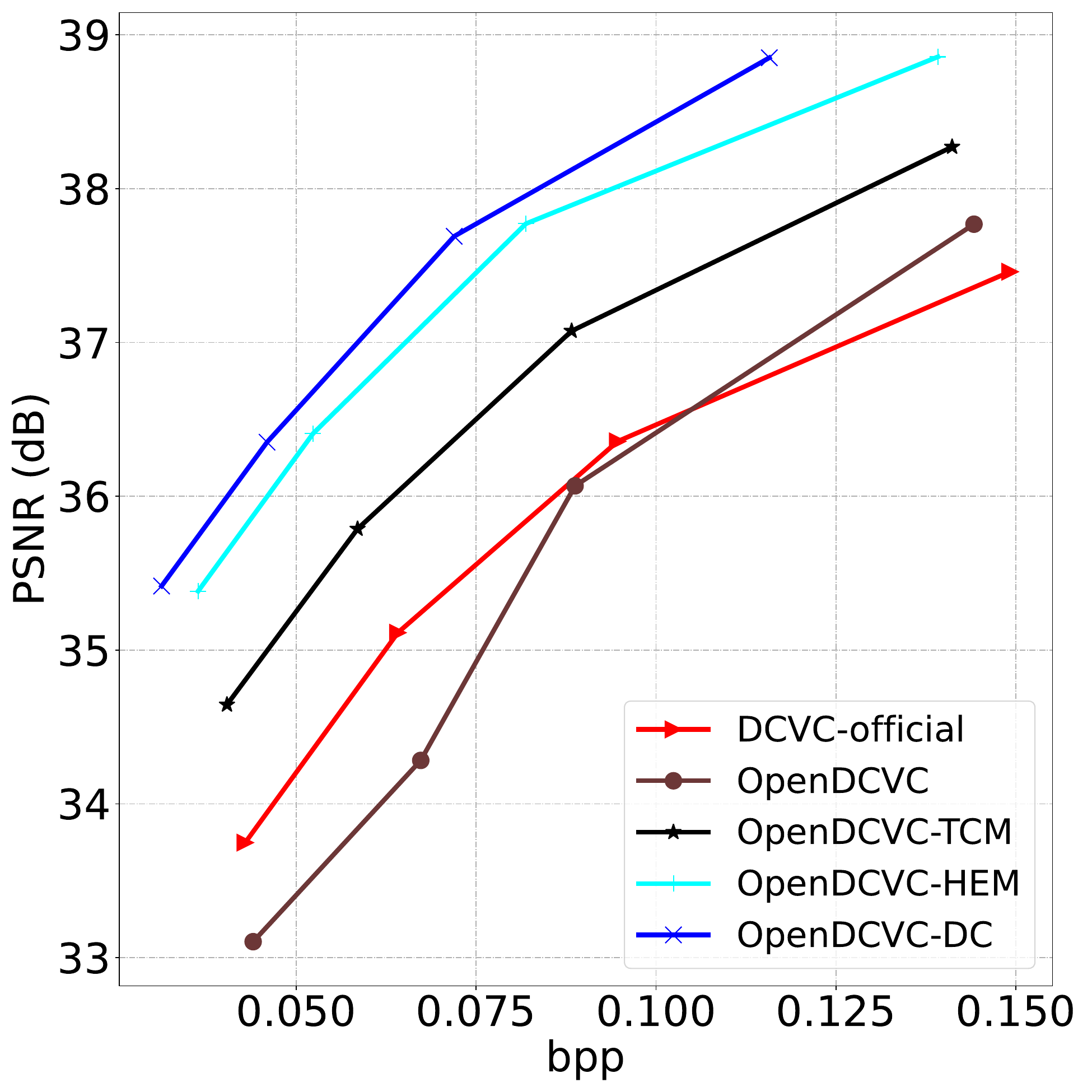}
    \caption{MCL-JCV}
    \label{subfig:Tecnick}
\end{subfigure}
\caption{\textbf{R-D curves of various methods. }{\it Please zoom in for more details}.} 
\label{fig:rd_fig} 
\end{figure*}

\subsection{Rate-Distortion Performance Evaluation}
The R-D performance of each algorithm in our library is illustrated in Fig.~\ref{fig:rd_fig}. Among all methods, OpenDCVC-DC achieves the best overall R-D performance across the evaluated datasets. Notably, our OpenDCVC implementation slightly outperforms the official DCVC results on high-rate points, demonstrating the effectiveness of our open-source optimizations.

For a quantitative comparison, Tab.~\ref{tab:rate-accuracy} presents the BD-Rate results of various methods relative to the official DCVC model results. OpenDCVC-DC shows substantial BD-Rate reductions on all datasets, with an average BD-Rate improvement of $-59.93\%$, confirming its superiority in compression efficiency. Other variants, such as OpenDCVC-TCM and OpenDCVC-HEM, also provide significant gains compared to the baseline.

\begin{table*}[htbp]
\centering
\caption{Comparison of various methods}
\label{tab:rate-accuracy}
\begin{threeparttable}
\begin{tabular}{l|c|c|c|c|c|c|c}
\hline\hline
\multirow{2}{*}{\textbf{Method}} & \multicolumn{4}{c|}{\textbf{BD-Rate (\%)}} & \multirow{2}{*}{\textbf{\makecell[c]{Model \\ Parameters}}} & \multirow{2}{*}{\textbf{\makecell[c]{Inference \\ Time}}} & \multirow{2}{*}{\textbf{\makecell[c]{GPU Memory \\ Occupancy}}} \\
\cline{2-5}
& HEVC-B & UVG & MCL-JCV & Average & & & \\ \hline
DCVC-official & 0\%& 0\%&0\% &0\% & 7.94 M& 0.2615 s& 21.79 GB \\
OpenDCVC &-10.60\% &-6.35\% & 10.40\%&-2.18\%& 7.94 M& 0.2620 s& 21.80 GB \\
OpenDCVC-TCM &-42.35\% &-46.70\% & -27.11\%& -38.72\%&10.70 M&0.3070 s & 5.67 GB \\
OpenDCVC-HEM &-56.39\% & -59.75\%&  -46.94\%& -54.36\%& 17.52 M& 0.3458 s & 4.74 GB\\
OpenDCVC-DC & -61.56\%& -65.49\%& -52.74\%& -59.93\%& 19.77 M & 0.5255 s&7.78 GB \\\hline\hline
\end{tabular}
\begin{tablenotes} 
    \item Testing Conditions: 1 $\times$ Nvidia L40S GPU, AMD EPYC 9554 CPU, 384GB RAM. Time is tested on all three 1080p test datasets.
\end{tablenotes}
\end{threeparttable}
\end{table*}

\subsection{Complexity Evaluation}
Tab.~\ref{tab:rate-accuracy} also summarizes the model parameters, inference time (excluding entropy coding), and peak GPU memory occupancy for each method. While OpenDCVC-DC provides the best compression performance, it requires more model parameters and slightly higher inference time compared to the baseline. However, memory consumption is significantly reduced in some variants (e.g., OpenDCVC-TCM and OpenDCVC-HEM), making them attractive choices for resource-constrained environments.

\section{Conclusion}
In this paper, we release \textbf{OpenDCVCs}, an open-source implementation providing a comprehensive, training-ready version of the DCVC series video codecs. The implementation is designed for ease of use and extensibility, supporting both training and evaluation. We introduce the modifications for training readiness and provide detailed benchmarking results, analyzing rate-distortion performance, inference time, and memory usage. We hope that OpenDCVCs will accelerate research and development in learned video compression by offering a solid, transparent codebase, and we will continue working on incorporating more algorithms.

{
\small
\bibliographystyle{IEEEtran}
\bibliography{sample-base}

\begin{thebibliography}{10}
\providecommand{\url}[1]{#1}
\csname url@samestyle\endcsname
\providecommand{\newblock}{\relax}
\providecommand{\bibinfo}[2]{#2}
\providecommand{\BIBentrySTDinterwordspacing}{\spaceskip=0pt\relax}
\providecommand{\BIBentryALTinterwordstretchfactor}{4}
\providecommand{\BIBentryALTinterwordspacing}{\spaceskip=\fontdimen2\font plus
\BIBentryALTinterwordstretchfactor\fontdimen3\font minus \fontdimen4\font\relax}
\providecommand{\BIBforeignlanguage}[2]{{%
\expandafter\ifx\csname l@#1\endcsname\relax
\typeout{** WARNING: IEEEtran.bst: No hyphenation pattern has been}%
\typeout{** loaded for the language `#1'. Using the pattern for}%
\typeout{** the default language instead.}%
\else
\language=\csname l@#1\endcsname
\fi
#2}}
\providecommand{\BIBdecl}{\relax}
\BIBdecl

\bibitem{lu2019dvc}
G.~Lu, W.~Ouyang, D.~Xu, X.~Zhang, C.~Cai, and Z.~Gao, ``Dvc: An end-to-end deep video compression framework,'' in \emph{Proceedings of the IEEE/CVF conference on computer vision and pattern recognition}, 2019, pp. 11\,006--11\,015.

\bibitem{li2021deep}
J.~Li, B.~Li, and Y.~Lu, ``Deep contextual video compression,'' \emph{Advances in Neural Information Processing Systems}, vol.~34, pp. 18\,114--18\,125, 2021.

\bibitem{sheng2022temporal}
X.~Sheng, J.~Li, B.~Li, L.~Li, D.~Liu, and Y.~Lu, ``Temporal context mining for learned video compression,'' \emph{IEEE Transactions on Multimedia}, vol.~25, pp. 7311--7322, 2022.

\bibitem{li2022hybrid}
J.~Li, B.~Li, and Y.~Lu, ``Hybrid spatial-temporal entropy modelling for neural video compression,'' \emph{Proceedings of the ACM International Conference on Multimedia}, pp. 1503--1511, 2022.

\bibitem{li2023neural}
------, ``Neural video compression with diverse contexts,'' \emph{Proceedings of the IEEE/CVF Conference on Computer Vision and Pattern Recognition}, pp. 22\,616--22\,626, 2023.

\bibitem{li2024neural}
------, ``Neural video compression with feature modulation,'' \emph{Proceedings of the IEEE/CVF Conference on Computer Vision and Pattern Recognition}, pp. 26\,099--26\,108, 2024.

\bibitem{agustsson2020scale}
E.~Agustsson, D.~Minnen, N.~Johnston, J.~Balle, S.~J. Hwang, and G.~Toderici, ``Scale-space flow for end-to-end optimized video compression,'' in \emph{Proceedings of the IEEE/CVF Conference on Computer Vision and Pattern Recognition}, 2020, pp. 8503--8512.

\bibitem{yang2020opendvc}
R.~Yang, L.~Van~Gool, and R.~Timofte, ``Opendvc: An open source implementation of the dvc video compression method,'' \emph{arXiv preprint arXiv:2006.15862}, 2020.

\bibitem{hu2021fvc}
Z.~Hu, G.~Lu, and D.~Xu, ``Fvc: A new framework towards deep video compression in feature space,'' \emph{Proceedings of the IEEE/CVF Conference on Computer Vision and Pattern Recognition}, pp. 1502--1511, 2021.

\bibitem{gao2023opendmc}
W.~Gao, S.~Sun, H.~Zheng, Y.~Wu, H.~Ye, and Y.~Zhang, ``Opendmc: An open-source library and performance evaluation for deep-learning-based multi-frame compression,'' \emph{Proceedings of the ACM International Conference on Multimedia}, pp. 9685--9688, 2023.

\bibitem{zhang2021dvc}
S.~Zhang, M.~Mrak, L.~Herranz, M.~G. Blanch, S.~Wan, and F.~Yang, ``Dvc-p: Deep video compression with perceptual optimizations,'' \emph{Proceedings of the International Conference on Visual Communications and Image Processing}, pp. 1--5, 2021.

\bibitem{lin2020m}
J.~Lin, D.~Liu, H.~Li, and F.~Wu, ``M-lvc: Multiple frames prediction for learned video compression,'' in \emph{Proceedings of the IEEE/CVF conference on computer vision and pattern recognition}, 2020, pp. 3546--3554.

\bibitem{minnen2020channel}
D.~Minnen and S.~Singh, ``Channel-wise autoregressive entropy models for learned image compression,'' \emph{Proceedings of the IEEE International Conference on Image Processing}, pp. 3339--3343, 2020.

\bibitem{zhang2024theoretical}
Y.~Zhang, Z.~Duan, Y.~Huang, and F.~Zhu, ``Theoretical bound-guided hierarchical vae for neural image codecs,'' \emph{Proceedings of the IEEE International Conference on Multimedia and Expo}, 2024.

\bibitem{lu2020content}
G.~Lu, C.~Cai, X.~Zhang, L.~Chen, W.~Ouyang, D.~Xu, and Z.~Gao, ``Content adaptive and error propagation aware deep video compression,'' \emph{Proceedings of the European Conference on Computer Vision}, pp. 456--472, 2020.

\bibitem{xue2019video}
T.~Xue, B.~Chen, J.~Wu, D.~Wei, and W.~T. Freeman, ``Video enhancement with task-oriented flow,'' \emph{International Journal of Computer Vision}, vol. 127, no.~8, pp. 1106--1125, 2019.

\bibitem{jvetj1010}
J.~M. Boyce, K.~Suehring, X.~Li, and V.~Seregin, ``{JVET-J1010: JVET Common Test Conditions and Software Reference Configurations},'' Joint Video Experts Team (JVET) of ITU-T SG 16 WP 3 and ISO/IEC JTC 1/SC 29/WG 11, San Diego, US, JVET Document JVET-J1010, July 2018, 10th Meeting, 10--20 Apr. 2018.

\bibitem{uvg2021}
``Ultra video group test sequences,'' \url{http://ultravideo.cs.tut.fi}, 2021, online; accessed 12 April 2021.

\bibitem{wang2016mcl}
H.~Wang, W.~Gan, S.~Hu, J.~Y. Lin, L.~Jin, L.~Song, P.~Wang, I.~Katsavounidis, A.~Aaron, and C.-C.~J. Kuo, ``Mcl-jcv: a jnd-based h. 264/avc video quality assessment dataset,'' \emph{Proceedings of the IEEE International Conference on Image Processing}, pp. 1509--1513, 2016.

\end{thebibliography}
}

\end{document}